# Engineering Token Economy with System Modeling

Zixuan Zhang

**Acknowledgment**

This thesis will not be possible without the advice and guidance from Dr Michael Zargham, founder of BlockScience and Penn PhD alumnus in Network Science and Dr Victor Preciado, Raj and Neera Singh Professor in Network and Data Sciences.

**Abstract**

Cryptocurrencies and blockchain networks have attracted tremendous attention from their volatile price movements and the promise of decentralization. However, most projects run on business narratives with no way to test and verify their assumptions and promises about the future. The complex nature of system dynamics within networked economies has rendered it difficult to reason about the growth and evolution of these networks. This paper drew concepts from differential games, classical control engineering, and stochastic dynamical system to come up with a framework and example to model, simulate, and engineer networked token economies. A model on a generalized token economy is proposed where miners provide service to a platform in exchange for a cryptocurrency and users consume service from the platform. Simulations of this model allow us to observe outcomes of complex dynamics and reason about the evolution of the system. Speculative price movements and engineered block rewards were then experimented to observe their impact on system dynamics and network-level goals. The model presented is necessarily limited so we conclude by exploring those limitations and outlining future research directions.

**Table of Content**

# Introduction

## A Token Economy

The advent of Bitcoin ushered in a wave of blockchain projects with native currencies, distributed architecture, and cryptographic guarantees. Many projects are built on the business narrative of a new digital economy where one can provide service or product and get paid in cryptocurrencies that can be sold on exchanges around the world. Given the decentralized nature of most of these platforms, there is almost no external authority that governs the evolution of these networks. These self-sovereign networks with emergent properties are only controlled by the rules of the system and the human agents that interact with them. It has become increasingly important to design the right set of microscopic incentives that can achieve the desirable system-level behavior. With economic incentives, human decisions, and cryptographic proofs, the term cryptoeconomics is coined to describe the design and study of incentives and mechanisms in a blockchain network.

Many approach the design problem with theories from mechanism design and algorithmic game theory whose analyses focus on the equilibrium [1]. However, the real world is full of random disturbances, irrational behaviors, and deviation from the equilibrium which goes beyond the scope of those analyses. We draw inspiration from early military pursuer-and-evader differential games where the model is agnostic to the exact behavior of the players and derives proofs just from permissible actions and trajectories of the system [2]. In the context of designing a blockchain economic system, as protocol designers, we want to design a set of rules such that irrespective of exact agent behaviors, the system-level objectives are still preserved. Compared to other games and distributed control considered in the past [3, 4, 5], putting protocol design in the context of a game has the property of global information where every agent is aware of the actions and payoffs of other players.

Our Markovian dynamical system modeling approach for decision making and economics drew inspiration from Ole Peters and Alexander Adamou's work on ergodicity economics [6] and John Sterman's work on business dynamics [7]. In addition, a stochastic dynamical system model enables us to understand complex relationships within a system and observe the business level impact resulted from secondary or tertiary dynamics. Non-obvious business impacts that are difficult to conceptualize will become clear and obvious after one has witnessed the impact as an outcome of a dynamical system. Such an impact is often non-intuitive prior to the observation.

## Differential Games and Control

Differential games bridge many concepts in control theory with game theory. Control theory is foundational to design systems that are robust to environmental noise and system-failures [8]. In our previous paper [9], we have described an application of state-space model to the analysis of blockchain networks. In this thesis, we will further describe the process of using state-space model to control and optimize for desired properties from the perspective of a protocol designer. To fully harness the power of control theory, we first need to define the system-level objectives for the blockchain network which are often neglected in the business narratives. In addition, traditional control theory often deals with physical system over which designers have direct control [10]. In designing an economic network of human agents and incentives, designers at best have indirect control over the incentive structure with little control over the exact behavior.

Since a large part of the design depends on the system-level objective for control engineering principles to work, we should first define the system in its state variables, transition functions, system-level objectives and cost constraints. Common system-goals include steady appreciation of network token, low cost of service provided on the network, steady growth of network utilization and so on. We will then have the potential to apply Lyapunov-style arguments to bound the reachable states of the system through the bound on its inputs [11]. However, to do a full proof of min-max optimization with Lyapunov control over input and output requires much more rigorous analysis and much contextual knowledge. Protocol designers will also require the power to design a universe where they can design incentive structure and energy function such that it requires energy for users to go against the allowable option space. Traditional control engineering concept such as a barrier function can be applied in constraining the allowable actions and reachable space. While this is a very interesting area, this thesis will not focus on its proof and analysis but instead setting up a generalized token economy model that other more complex models and system can be extended from.

Another major difference of our model from classical control system model is the presence of a black box. We see this in price movement but also in agent behavior. Each individual follows their own control policies, not necessarily the ones that protocol designers hope for. This is where concepts from differential games come in. Agents follow their own policies as a function of their own beliefs of the system state and their own payout function. The collective behaviors of different agents result in sometimes unexpected system-level outcome. A differential game assumes an adversarial agent with a complete opposite payout function. The role of the designer is to design a set of rules and incentives such that the system-level goal can still be achieved irrespective of the exact behavior of the agents. In our model, we will treat these black boxes as random variables drawn from a reasonable distribution from which some meaning can be derived.

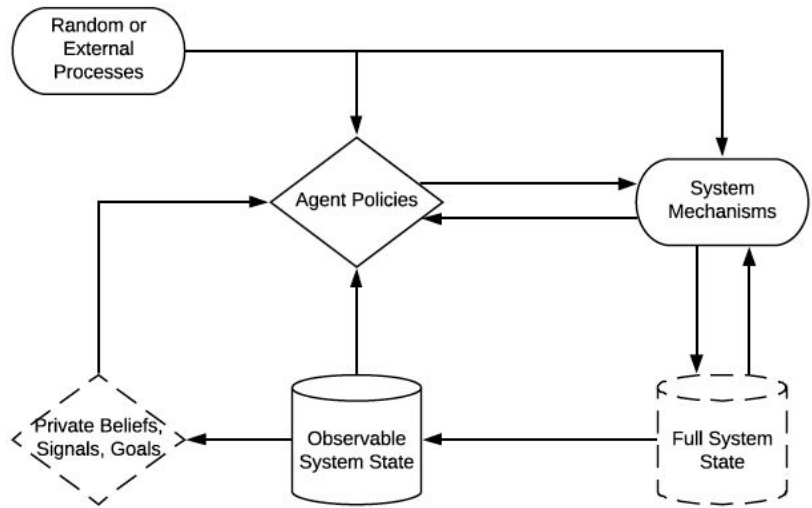

Figure 1. High-level control systems for agent-based networks.

As illustrated in the figure above, the exact agent beliefs and the full state of the system are much like black boxes. Random processes will be fed into the agent policies and system mechanism which collectively update the unobservable full system state. Agents in the system can then observe the state and form their beliefs, signals, and objectives that impact their policies in the next iteration. Despite the presence of these unobservable states and random processes, we can design allowable agent policies and system mechanisms to achieve certain observable outcome. This mix of white-box and black-box approach renders the token design process robust and innovative.

## Networked Economy

Another key aspect of a token economy is that it is a global networked economy from day one, from the physical layer of distributed machines that power the network to the agent level network of producers and consumers of services [9]. There are many interesting topics within the domain of network itself that are worth considering. Quite a few papers in network formation will prove relevant in modeling and understanding the growth of a network [12, 13]. In addition, the contact and interaction of agents on the network are also dependent on the network topology, similar to how epidemic spreads over different network topologies [14, 15]. Network topologies play an important role in network efficiency, growth, and even adoption.

However, as our thesis focuses on the economics of the token economy on a population level, the exact interaction over network topology is important but less relevant in the immediate term.

Many of the cryptoeconomic networks are still in their infancy for us to consider their topologies. In addition, much of the impact of network topology is reflected in network latency which at best can lead to lag in decision and processes. For example, mining nodes operating in a mining farm share close proximity with each other and result in reduced latency and oftentimes lead to gains in consensus power [16, 17]. In our population level analysis of a service based economy, the impact of network topology has been absorbed and reflected in the quality of service, in terms of up time, response time, and latency. From a population-level economy's perspective, it may appear to be a random noise in the system and hence, we will not include network topology in our modeling and analysis. Nonetheless, our model lays down the foundation for future work and enhancement in this area.

## Beyond Cryptocurrencies

Even though the problem that we are solving stems from designing cryptoeconomic network, there are many parallels and similarities between designing the rules to deploy resources in a blockchain network and allocating resources to build a network economy with venture capital. After all, the generous block rewards subsidizing a new token economy is very similar to the subsidies for new platform businesses provided by venture capital money, like Airbnb, Uber, and Lime. Perhaps they know how to allocate resources and design incentives to achieve system-level objectives for their networked economy.

From an economic research perspective, our approach has well addressed the Lucas critique in macroeconomics. Lucas critique argues that it is naive to solely rely on past data and correlation and neglect fundamental micro-foundations in designing economic policies and making economic decisions [18]. It is worth noting that our approach to engineering economies address this critique head-on. We start from agent-level incentives and policies with the system state as an input. We will then run simulation with state updates to help further refine the agent policies and shed light on system level decisions. This combination of microeconomic incentives and macroeconomic outcome connected through networked system dynamics underpins the robustness of our model and approach.

# Model Setup

To facilitate proper design and engineering of a token economy, we need to first define its internal states and transition dynamics. First of all, let us first define a token economy and how it might be different from a traditional platform economy despite their similarities. Next, we will highlight the assumptions that we will be using in our model. Lastly, we will define the minimal set of internal states and their dynamics.

Traditional networked economies have been characterized by a centralized platform where users and producers can exchange services on this platform. The value of such a platform is often considered to be proportional to the square of the number of active users by Metcalfe's Law [19]. While Bitcoin and other dominant blockchain platforms today have been criticized for their high energy usage and wasteful computation with its Proof of Work, there are new token economies introducing new models where the work done by miners can be useful and dubbed Proof of Useful Work. The work that miners have done is both useful to the users and useful in maintaining consensus and network security. New networked economies such as incentivized peer-to-peer file storage network and video streaming network are just two promising examples [20, 21].

For most of these token economies, platform providers and service producers are tightly coupled even though they have plans to allow for third-party providers on their network to provide differentiated services. Apart from that, there are protocol designers and organizations designing state update mechanisms and policies within the system but not actively participating in the economy. As such, in our models below, we will combine platform providers and service producers into miners and we will put ourselves in the shoes of protocol designers with design freedom over state update mechanisms.

## Model Assumptions

To help formulate the model for a token economy, we will first define the assumptions that we are making. After all, the quality of a model is only as good as its assumptions.

Assumption 1:

The model considers each miner and user homogeneous with unit service capacity similar to work in mean field games [22]. A large miner in real life will just be an aggregation of many unit miners in the model. Similarly, a user with a lot of demand will be a combination of many unit users. Each unit miner and unit user supply and demand one unit of service which is defined by the users of the model. This abstracts away individual idiosyncratic demand and supply from users and miners. Each user demands one unit of service and each miner can only provide one.

Assumption 2:

We assume a perfectly competitive market where miners are users are both price takers, given the open nature of a token economy [23]. Users will consume inversely proportional the unit price of service denominated in fiat currency. Similarly for miners, if mining is very profitable, miners will provide as much service as possible to earn block rewards. Furthermore, we assume the service that miners provide is a commodity, meaning that it does not matter who produces the unit service. This is in alignment with the assumption of a perfectly competitive market where products are homogeneous with no differentiation [24].

## Model Scope

With these assumptions in mind, we can create a state-space representation of the system. We aim to define the minimal set of internal states that can capture different aspects of the system. The system can pause and restart any time if we keep track of all the state variables. We will also define TOKEN, or TOK in short, as the native cryptocurrency to the network. There are two subsystems within our token economy. The first is with regard to the flow of TOK and the second is related to the service provided on the network. The two subsystems are connected with three important signals, miner's profitability, price of service on the platform, and the price of the token itself. We also assume a discrete time system and states are only updated after a new block is produced. Note that it can also be extended into an event-based discrete time system.

If we first look at where TOKs are held, they can be in the hands of miners, users, and a liquid pool. We are calling them TOK miner, TOK user, and TOK liquid. In many cases, miners are required to pose a collateral requirement in TOK to participate in the network and block rewards are minted directly to miners' account [25]. Miners allocate a portion of their TOK to participate in the mining process to earn more block rewards and may sell some of them to the liquid pool when the price is high. Similarly, users buy TOKs from the liquid pool to purchase service from miners or they can speculate and trade in the market. Here, we can encode different user and miner population profiles in which they have different strategies and thresholds with regard to trading TOK [26]. The flow of TOK among different holding parties within the system can be illustrated with Figure 2 below.

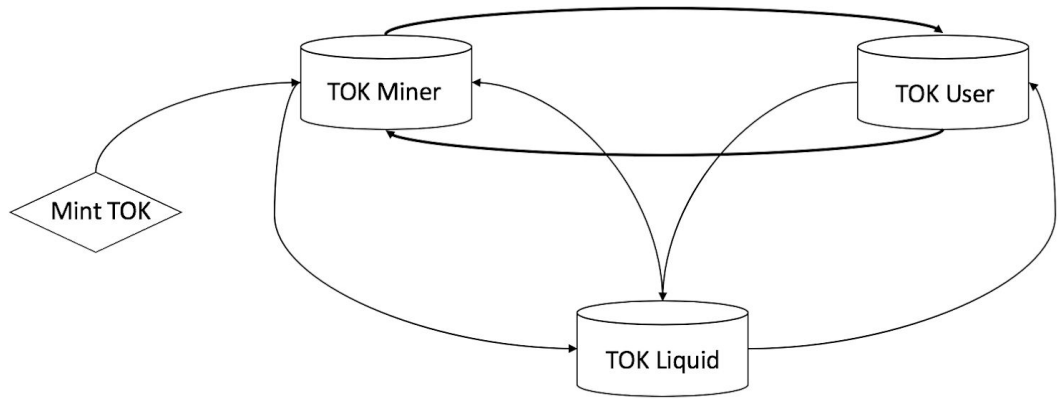

Figure 2. Flow diagram of TOK holdings.

The change in total TOK owned by miners is constantly affected by the inflow of block rewards and transaction fee paid by the users. Depending on the price of TOK at t, miner population will sell a fraction of its share of TOK to the liquid market. Similarly for users, rational users may hoard TOK when the price is low and attempt to sell them when the price is high. Users who are using the service on demand will purchase TOK when the service price is attractive and users will have to pay for a transaction fee when service is consumed. Miners and users buy from and sell to the liquid market which represents the total tradeable TOK available in the market. Fundamentally, network mints TOK in exchange for the service that miners provide, even though in an inflationary way.

Given our efficient market assumption, we have decided to leave out the modeling of TOK holdings entirely. For one, it is one subsystem in and of itself with its own dynamics about price movement relative to the amount of TOK in the liquid pool and trading activities. In addition, with the efficient market assumption, we can assume that all the trading volume and liquidity has been embedded in the price. After all, both miners and users are just price takers. Lastly, within TOK user, there is also a group of speculative investors with a different set of dynamics. The dynamics of speculative investors catalyzing technological innovation is a very interesting topic but yet another subsystem that can be built as an extension to our model.

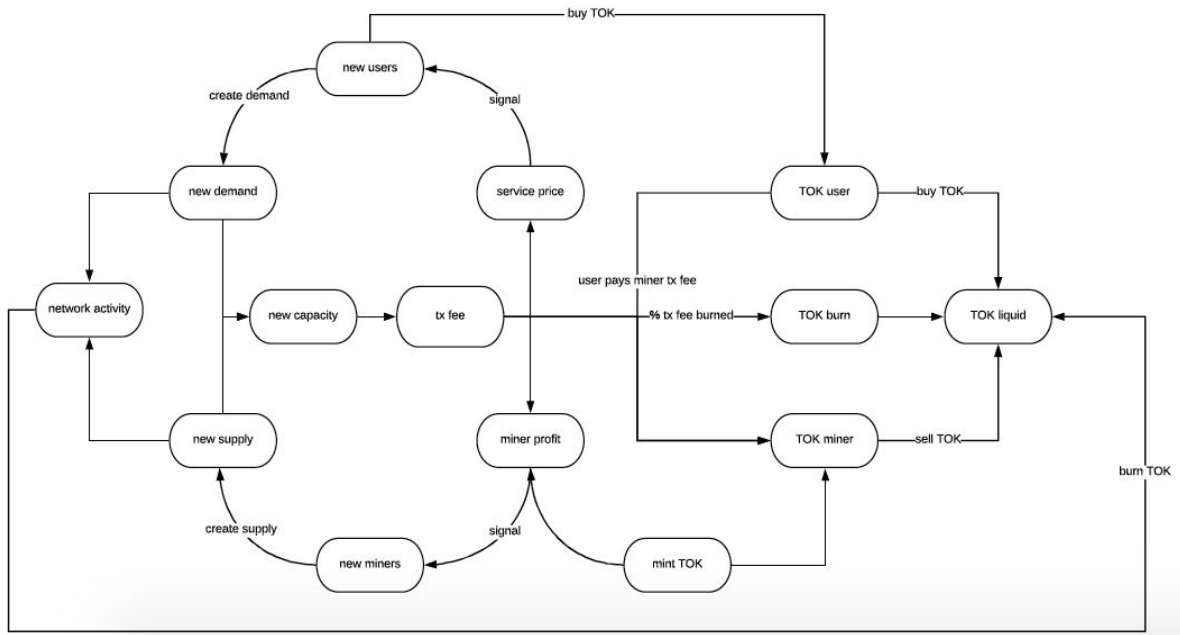

Figure 3. Feedback mechanism in a token economy.

Figure 3 presents a more complete picture of feedback mechanism in a token economy. There are two main subsystems, one with the flow in TOK holdings and the other with the network service provided. The efficient market assumption allows us to treat price of TOK as a signal that encapsulates the flow in TOK holdings. Any attempts to take flows in TOK holdings into account at this point will over-complicate the model and introduce arbitrary dynamics. Nonetheless, it is worth considering and building a subsystem on top of that as an extension to our current model.

Hence, the scope of our model is restricted to an efficient token economy where miners and users provide and consume commodity services on the platform as price takers.

## State Variables

Given the above scope and assumptions, we will now define state variables for the token economy system that we would like to study.

**S(t):** supply of unit service at time t
**D(t):** demand of unit service at time t
**Q(t):** transacted service at time t
**P(t):** price of a unit service at time t
**R(t):** miner profitability at time t

**TOK(t)**: price of a TOK at time t
**B(t):** block rewards released at time t
**C(t):** cost of unit service at time t

There are also intermediary variables are relevant in state updates.

**ΔS(t):** arrival of new unit service supply at time t
**ΔD(t):** arrival of new unit service demand at time t
**V(t):** TOK earned per unit fiat invested in the system at time t
**1/V(t):** intrinsic value of TOK at time t
**U(t):** speculative value of TOK at time t
**W(t):** amount of TOK left in block rewards pool at time t
**M(t):** total amount of tokens that has been released as block rewards by time t
**KPI(t):** index on the network's progress to achieve its goal at time t

It is also worth noting that the price of TOK is not just dependent on the internal states of the system as TOK is openly traded on secondary markets from day one. Price of TOK will have significant impact on agent incentives within the network. It is also subject to influences from secondary market dynamics, speculation, and sentiments that may or may not correlate with the actual activity on the network itself.

## System Dynamics

We will first model supply and demand as an arrival-departure stochastic dynamical system. Miner profitability and unit service price are the two driving signals for two main feedback loops in the systems. Miner profitability represents how much the network is going to pay for the service that miners provide. A high miner profit will bring in more miners and supply of service. Similarly, unit service price represents how much the network is willing to accept in exchange for the service capacity it provides. When mining is extremely profitable, we can expect service price to be small and supply outweighs demand. A low unit service price will bring in more users and more service demand.

Hence, we are modeling the arrivals of new service supplied and new service demanded as two Poisson processes.

**ΔS(t) ~ Po(λs(t))**
**ΔD(t) ~ Po(λd(t))**

The mean of two Poisson processes are dependent on the unit service price and miner profitability. More concretely, **λs(t)** should expand when mining is very profitable and contract when it is less so. Similarly, **λd(t)** will expand when the price of a unit service is low and contract when the price of a unit service is high.

**λs(t) ~ λs(t-1) × (R(t) / R(t-1) )**
**λd(t) ~ λd(t-1) × (P(t-1) / P(t))**

We further define two constants as the departure rates for service supplied and demanded.
**Xs:** departure of unit service supply at every time step
**Xd:** departure of unit service demand at every time step

Having these two constants simplifies the dynamics without losing much of its meaning. In the case where supply is increasing quickly, we can consider service departure as departing and then immediately arriving again. This can be accounted for by a very positive **ΔS(t)**. The same can be said about departure in demand.

**S(t) = S(t-1) + ΔS(t) - Xs**
**D(t) = D(t-1) + ΔD(t) - Xd**

From **S(t)** and **D(t)**, we can define **P(t)** and **Q(t)**, which are the price of service and the amount of service consumed and transacted in every round.

**P(t) ~ D(t) / S(t)**

Given our efficient market assumption, price is simply set by relative strength in demand and supply. When demand is greater than supply, a higher price is expected and vice versa. This price model, despite being very simple, captures much of the dynamics in the system. It can be further expanded to include momentum, user valuation, and other factors that will affect service price.

**Q(t) ~ min(D(t), S(t))**

**Q(t)** is the amount of service transacted on the network at time t. It is at most the minimum between net new demand and supply. After all, no transactions will take place with unmet demand or supply. This can be made more realistic with a slippage later since not all matching supply and demand can find each other in the market. Nonetheless, this has been taken care of by the efficient market assumption.

We can now define the dynamics for the cost of providing a unit service, miner profitability, and price of TOK in our system.

We will first model **C(t)** as a stochastic process that is noisy but it is neither diverging or converging, similar to sampling from a normal distribution with momentum.

$$C(t) = \alpha C(t-1) + (1-\alpha)N(\mu, \sigma)$$

Then, we can define a signal **V(t)** which is the revenue per unit spend signal for miners from all the aforementioned state variables.

$$V(t) \sim (P(t) \times Q(t) + B(t)) / (C(t) \times Q(t))$$

**P(t) × Q(t)** is how much miners earn from providing the service in terms of TOK and **B(t)** is the subsidy that the network provides in TOK at time t. This represents how much tokens are miner earning for the service they provide. Multiplying **V(t)** which is in TOK/FIAT by the price of token at time t, **TOK(t)**, we get miner profitability in a unitless denomination. The system also cares about the inverse of **V(t)**, which is **1/V(t)**. This represents the intrinsic value of the token as it measures the value of service provided per unit of token in FIAT/TOK.

$$R(t) = V(t) \times TOK(t)$$

However, determining **TOK(t)** is not easy. The secondary market price is a speculative estimator of future **1/V(t)**. Market decouples from current state of **1/V(t)**, because speculation is estimating future **1/V(t)** in an effort to create returns. Hence, it makes sense to model price of TOK as a convex combination of its intrinsic and speculative value. This is in line with the general asset pricing framework that the price of an asset can be attributed to its fundamental and speculative value [27, 28, 29]. Speculative value **U(t)** captures momentum in price movement with a naive projection. In the equation below, **γ** is treated as a constant but it can also be a randomized value that can randomize the composition of the mixture model.

$$TOK(t) = \gamma \times (1 / V(t)) + (1 - \gamma) \times U(t)$$

Lastly, TOK issued in block rewards is usually set by a predetermined release schedule that follows an exponential decay rate. However, we can consider a more abstract and generalized version of block rewards issuance by introducing the concept of a **KPI(t)**. Most traditional exponential decay block rewards scheme with a pre-determined release schedule is effectively treating the block time as the KPI. However, if we define the release of block reward as a function of the rate of change in achieving the KPI, we can then create direct incentives based on what the network desires.

**M(t) = $\sum_{i=0}^{t}$ B(i):** total amount of tokens that has been released as block rewards by time t

**W(t) = W(0) - $\sum_{i=0}^{t-1}$ B(i) = W(0) - M(t - 1):** amount of TOK left in block rewards reserve

**ΔKPI(t) = KPI(t) / KPI(t-1):** change in KPI

**B(t) ~ ε × ΔKPI(t - 1) × W(t - 1)**

Subsidy can be given out as a function of the change in **KPI(t)** and the amount of TOK left in the rewards pool. This set of subsidies **{B(0), B(1), B(2), ..., B(t)}** can be the set of control policies that protocol designers can control to bootstrap the network to some target with some initial capital. This KPI can be as simple as the cumulative service transacted on the network over some period of time. Figure 4 below summarizes the system dynamics outlined above.

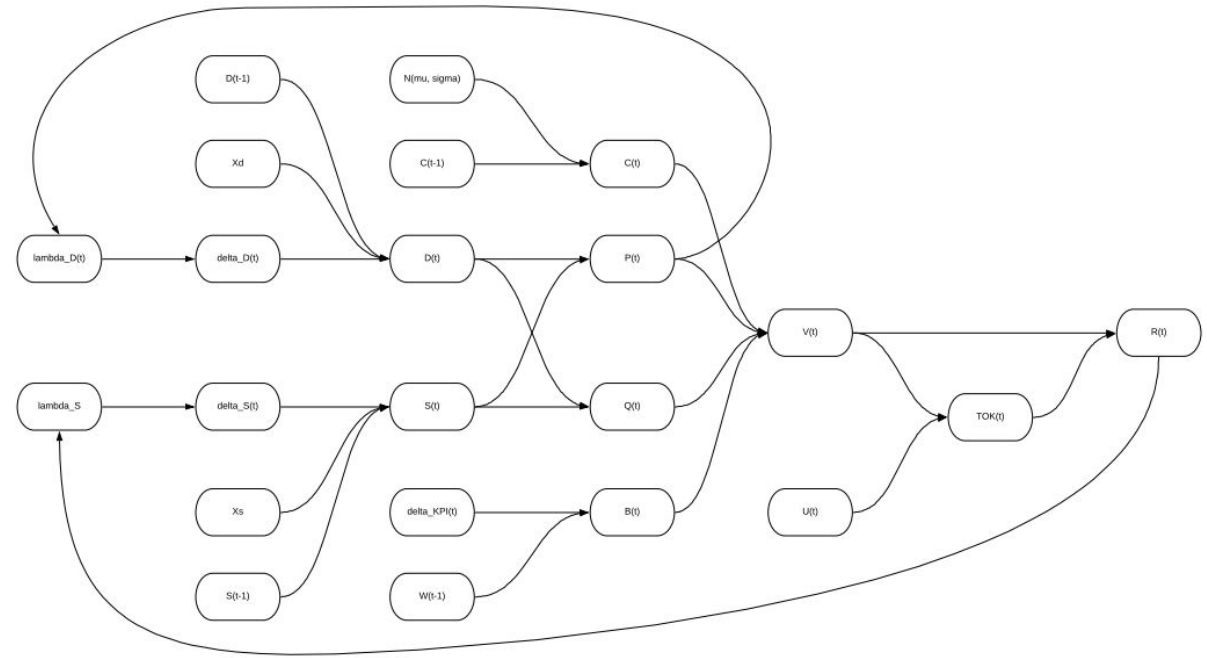

Figure 4. System dynamics at every round of update.

# Simulation and Evaluation

We are leveraging a tool, cadCAD, which stands for Computer Aided Design of Complex Agent-based Decision System, developed by Dr Zargham and his team at BlockScience. It is a Python differential game engine that assists in the process of designing, testing, and validating complex systems through simulation. We have encoded the dynamics of different agents and subsystems in every discrete time step into the system.

To facilitate evaluation, we want to outline a few metrics that we can use to define the success of our network. As a token-enabled platform economy, we would first want the platform to grow in adoption and in the underlying value of the token. In other words, we want **Q(t)**, **V(t)**, and **TOK(t)** to grow over time. We will also look at the distribution over the total aggregated growth of **Q(t)** in our model. In addition, we want a relatively low volatility in the price of the service provided **P(t) × TOK(t)** and hence we will look at the variance which is its mean squared error from its mean as a volatility metric. Too volatile a unit service price may render the platform service less attractive to potential users.

We will first come up with a baseline simulation in which the release of block rewards follows a traditional exponential decay with a predetermined half-life on a similar normalized scale to Bitcoin. We will then run Monte Carlo simulations for the baseline setup and plot summary statistics of the metrics that we care about. Next, we will experiment with different block rewards subsidy regime to show that it can engineered to improve macroeconomic outcomes.

## Baseline Simulation

As a baseline simulation, we are treating timestep as the KPI of the system. The concept of a KPI that the system is tracking can be applied to the timestep itself, demonstrating the generality of our model. The most common policy for block rewards is a halving schedule where the block rewards half after some period of time. Our baseline model will adopt a continuous exponential decay as the block reward release schedule, treating timestep as the KPI. To determine what is a good decay rate, we are looking at Bitcoin as the benchmark for our generalized token TOK. Note that almost every cryptocurrency follows a different schedule. Even in the case of Bitcoin, its block reward release function follows a step decay function with a halving period of 4 years, introducing arbitrary shocks into the system [30]. Hence, the goal for this comparison is to determine a comparable scale for the block reward schedule, rather than fitting for an estimate. Instead of simulating on the block by block level, we chose to simulate on a weekly basis. A week-by-week level makes it high-level enough to reason about the numbers and granular enough to experiment with policy changes.

| Currency | Total Supply | Initial Block Reward (weekly) | Half Life (weeks) |
|---|---|---|---|
| BTC | 21,000,000 | 50,400 | 208 (4 years) |
| TOK | 10,000,000 | 26,624 | 260 (5 years) |

Table 1. BTC and TOK release schedule comparison.

The decay rate can be back-calculated with the following formula derived from the sum of a geometric series.

$$Total\ Supply\ =\ \frac{Initial\ Block\ Reward}{1 - Decay\ Rate}$$

$$\tfrac{1}{2} Initial\ Block\ Reward\ =\ Initial\ Block\ Reward\ \times\ Decay\ Rate^{Half\ Life}$$

$$Decay\ Rate^{Half\ Life}\ =\ \tfrac{1}{2}$$

Hence, the initial block reward is 26624.00 and the decay rate should be 0.99734. We will run Monte Carlo simulations for 100 times for each configuration for 1040 weeks or 20 years to evaluate the model. For the baseline simulation, here is the block rewards release chart.

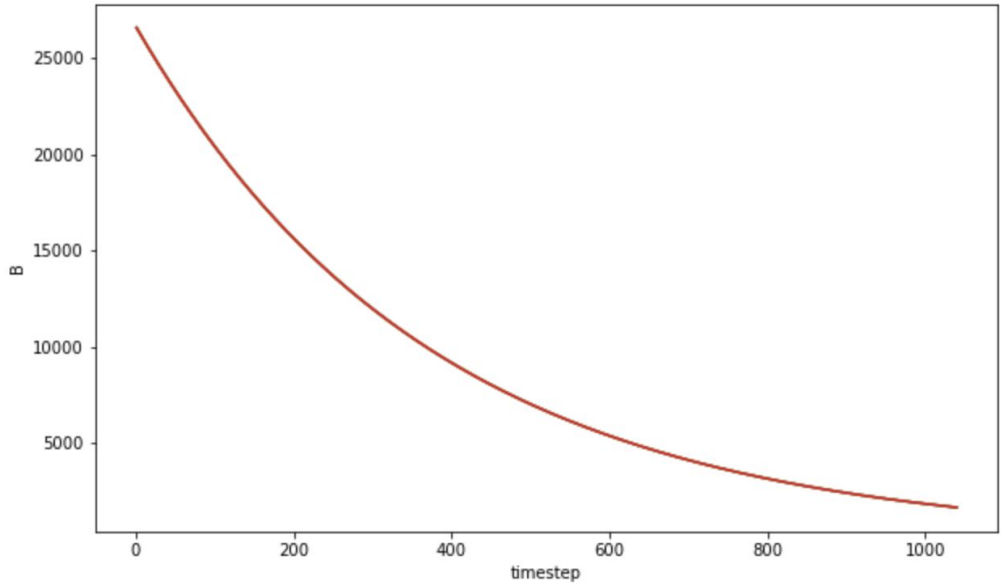

Figure 5. Baseline model open-loop deterministic block rewards B(t).

Next, we would like to define the following metrics that we care about as a system.

1. Growth trajectory in **1/V(t)** of unit FIAT/TOK.
2. High aggregated growth in **Q(t)** which measures the usage of the system and approximates network value.
3. Low variance on **P(t) × TOK(t)** as the price of the service in fiat unit.

The following are the charts and statistics of a particular simulation of the baseline model over 100 Monte Carlo runs.

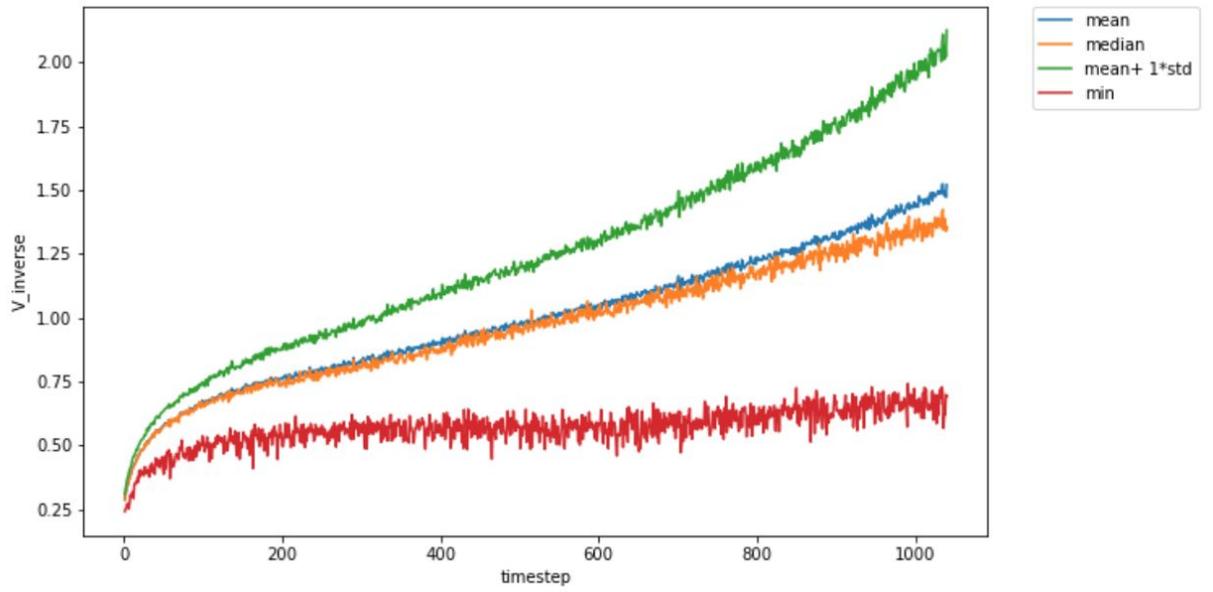

(a)

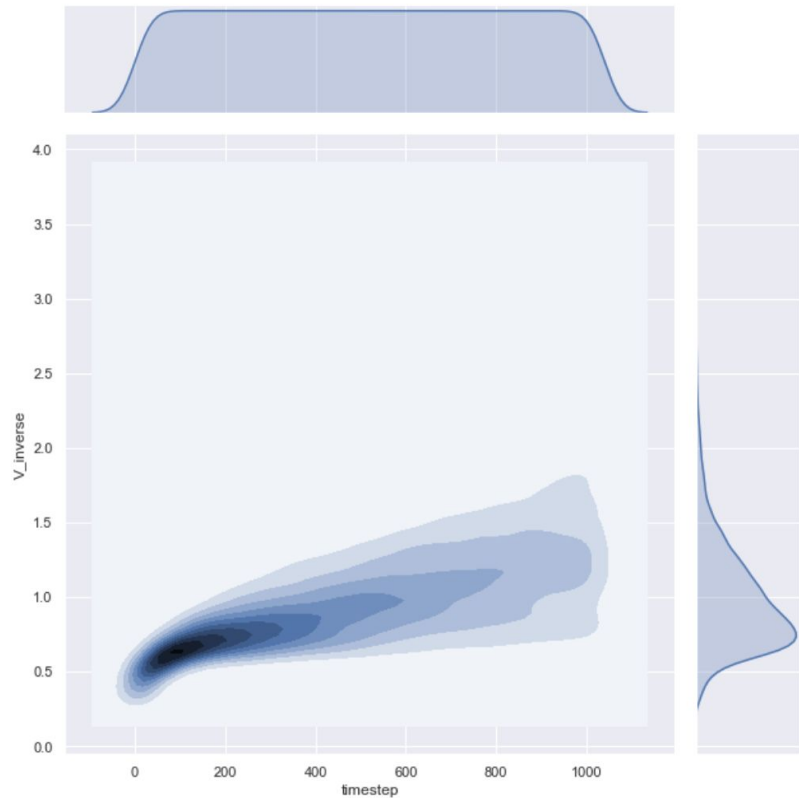

(b)

Figure 6. (a) Summary statistics and (b) distribution of $1/V(t)$ over 100 Monte Carlo runs for baseline simulation.

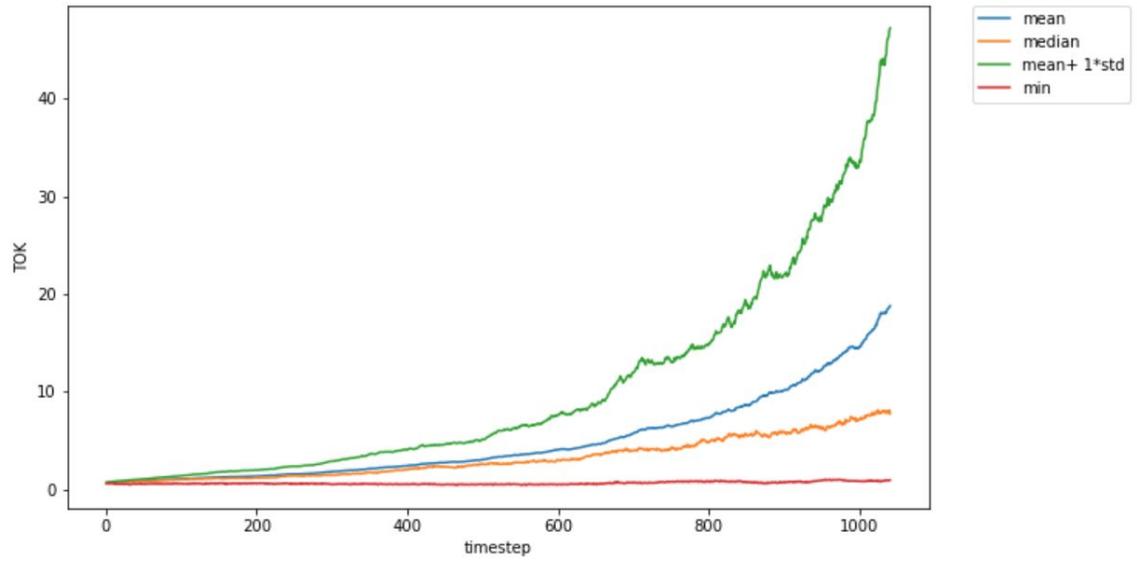

(a)

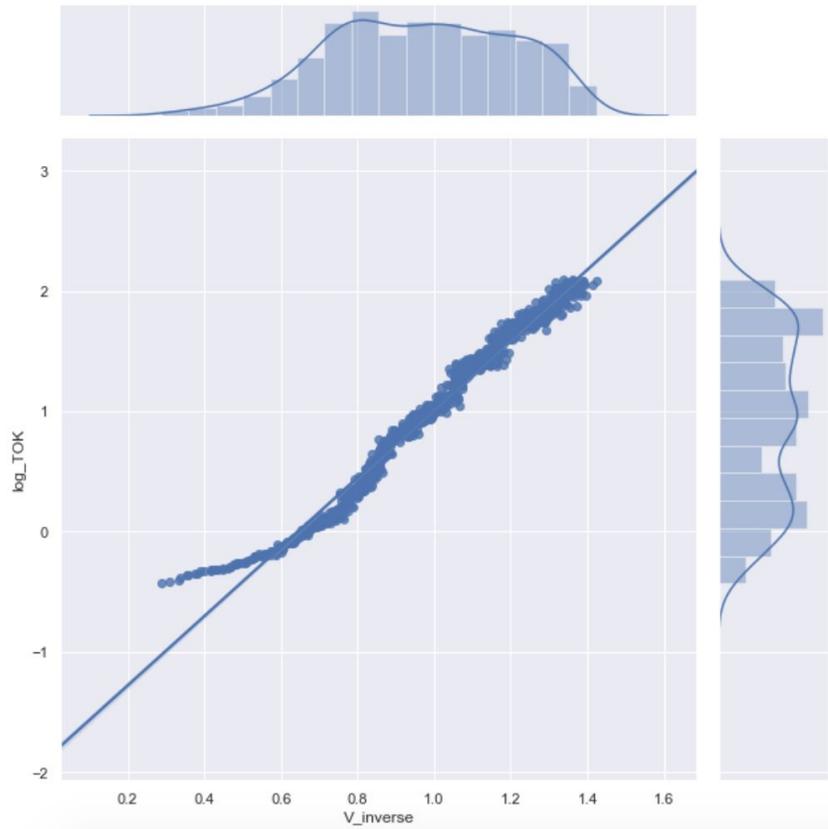

(b)

Figure 7. (a) Summary statistics of TOK(t) and (b) bivariate distribution of log(TOK(t)) and 1/V(t)

over 100 Monte Carlo runs for baseline simulation.

As we can see from Figure 6(a) and Figure 7(a), **1/V(t)** grows linearly and steadily whereas the token price on secondary market has increased exponentially. The mean token prices is at about a 20-time multiple of the mean intrinsic value, suggesting that the token price is largely driven by speculative value. This is further confirmed by the joint distribution of **log(TOK(t))** and **1/V(t)** in Figure 7(b). The log of **TOK(t)** is highly correlated with **1/V(t)**, especially so when the network has gained some intrinsic value so that speculators have something to speculate on.

However, the growth of the intrinsic value slows down rather quickly, as seen in Figure 6(a), while the token prices continue to grow in Figure 7(a). This can be attributed to how the secondary market prices are modeled in our system. It can also be explained by how secondary market is not trying to track intrinsic value but trying to track how much people consider the token to be worth in the future. This is rather common in the world of markets and finance where one hope to estimate the distribution and evolution of strategies by all players in the market.

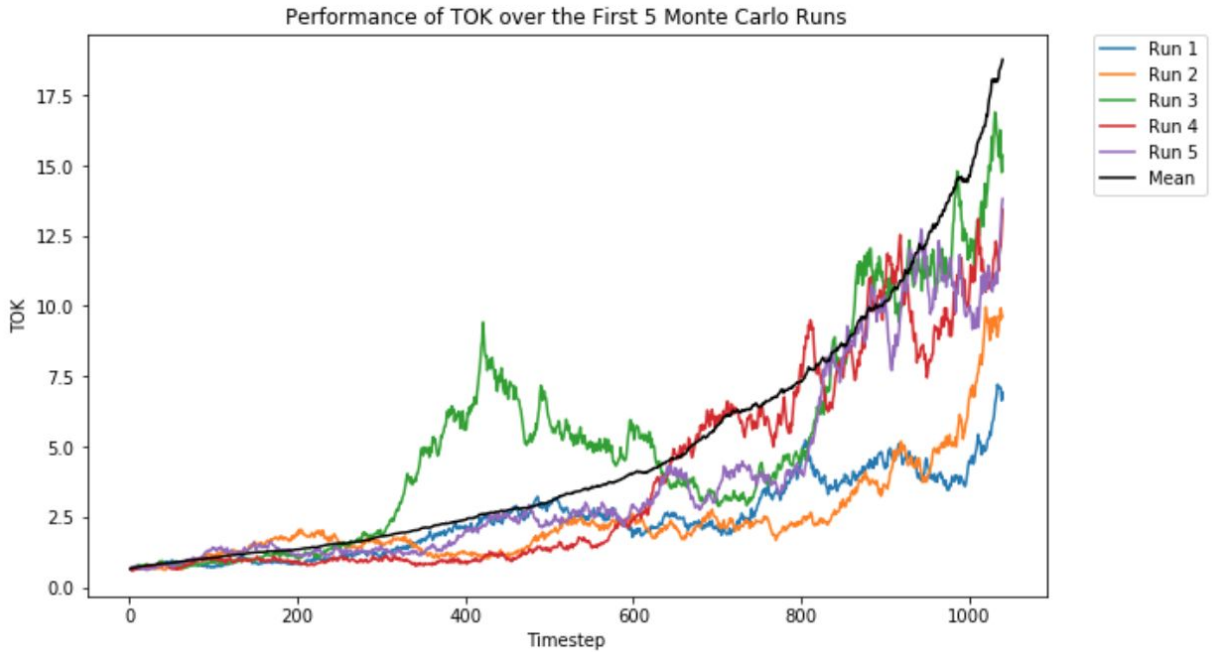

Figure 8. TOK movement for the first 5 Monte Carlo runs for baseline simulation.

A sanity check from Figure 8 shows that our price simulation which is primarily a Brownian motion with drift is largely in line with movement in cryptocurrency prices, especially in Run 3.

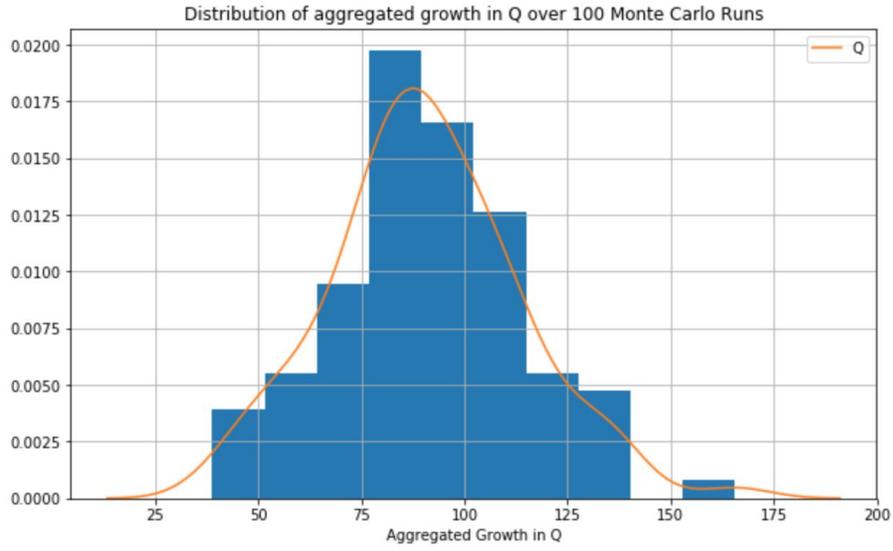

(a)

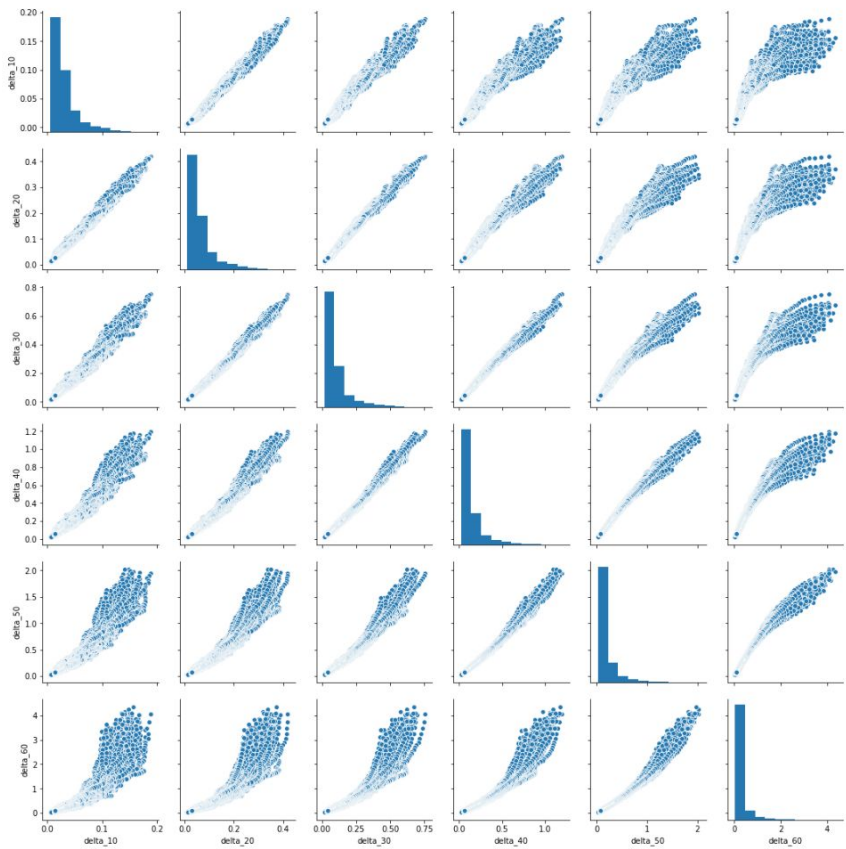

(b)

Figure 9. (a) Aggregated growth in Q(t) (b) pairwise distribution of growth in Q(t) over different time durations over 100 Monte Carlo runs for baseline simulation.

In Figure 9(a), we look at the aggregated growth in **Q(t)** which is the cumulative growth in transaction volume of the platform and we want to shift the overall distribution towards the right as much as possible. In Figure 9(b), we plot pairwise distribution of time averages of growth in **Q(t)**. We observed that the average growth rate across 10 timesteps is still highly correlated with the average growth rate across 20 and 30 timesteps. This makes sense because, after all, they share the same underlying data and short-term price movements tend to have higher correlation with each other. However, this correlation falls apart as the duration window increases, indicating that the system is not time-invariant. This is interpreted as a result of non-ergodicity because if the system is indeed ergodic, we should expect invariance on the temporal dimension. Since it is not observed, we infer that there is some element of non-ergodicity. Note that there is a clear upward trend structure in Figure 9(b), this can be explained by how the simulation is set up and **Q(t)** always increases with new arrivals of **S(t)** and **D(t)**. We can inject more entropy into the system by changing the initial values, arrival, and departure rates.

This is an important result to acknowledge because most economic systems and human activities are non-ergodic in nature whereas many Markovian systems and simulations are implicitly ergodic. Non-ergodicity means that the ensemble average across different parallel universes is independent of the time average of a particular trajectory of an individual. This is important because even if we can derive properties successfully from an ensemble average, there are no guarantees that a particular network economy's trajectory will be successful [31]. It will be interesting to further understand what contributes to the non-ergodicity in our system in future studies.

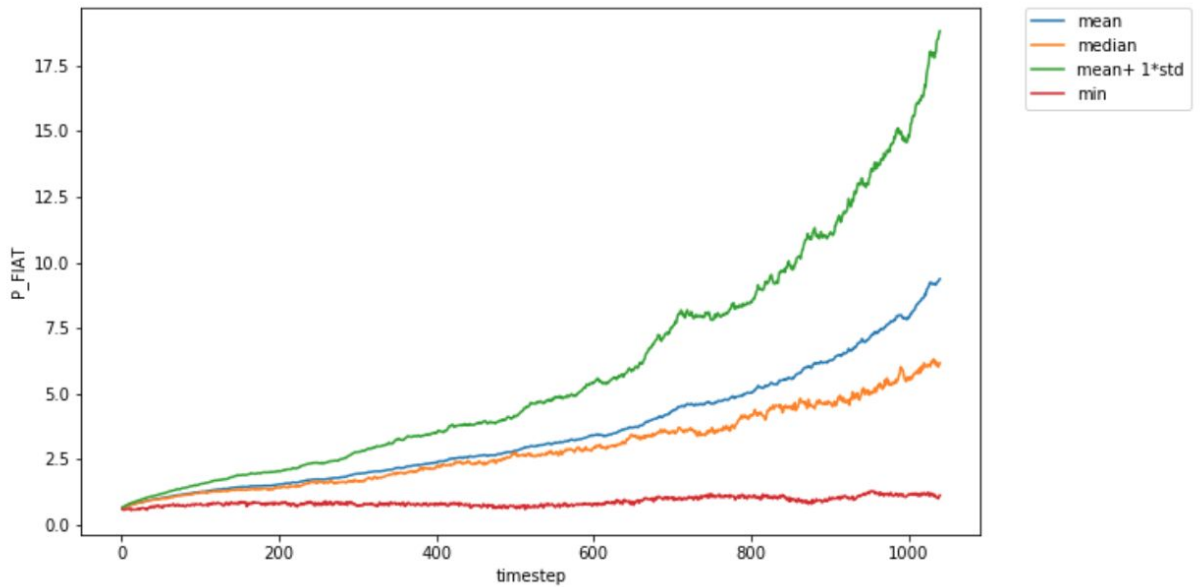

Figure 10. Summary of P(t) over 100 Monte Carlo runs for baseline simulation in fiat currency.

Figure 10 shows the movement of service price in fiat terms with a variance of 10.43 and a mean of 3.55. A steady and relatively low service price is desirable property that the network would like to pursue. Nonetheless, what is a low price for unit service depends on what the unit service itself is and it is up to the users to define.

## Speculative Price Influence

The price of the token has been the bridge between the token economy and the rest of the world economy. It is therefore interesting to look at the impact of price movement on system-level objectives and behavior. In addition, cryptocurrencies have always been notorious for their price volatility and speculative behaviors.

Speculative TOK prices play a dominant role in the **R(t)** which is the profitability level for miners. This will in turn have an impact on the supply of the service **S(t)** which will further impact service price **P(t)** and service transacted **Q(t)**. As a result, the higher the TOK prices, the greater the miner profitability, the higher the service supply, and the lower the service price. A cheaper service price creates greater service demand **D(t)** that consumes the service supply **S(t)**, leading to greater service transacted and hence even more profits for miners, creating a positive feedback loop as illustrated in Figure 11 below.

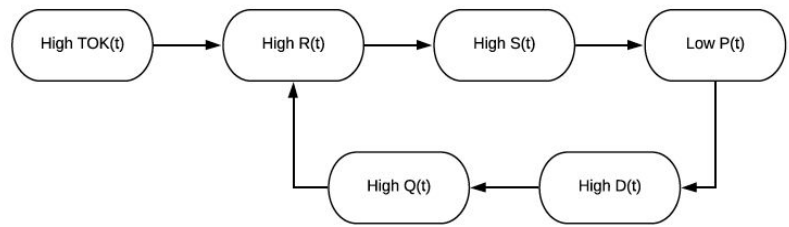

Figure 11. Positive feedback loop within token economy.

We now perform the following experiments with another 100 Monte Carlo simulations each at different upward drift in terms of TOK prices. In the baseline model, there is a 55% probability drift that TOK price will move up. We ran two separate experiments with 58% and 52% respectively.

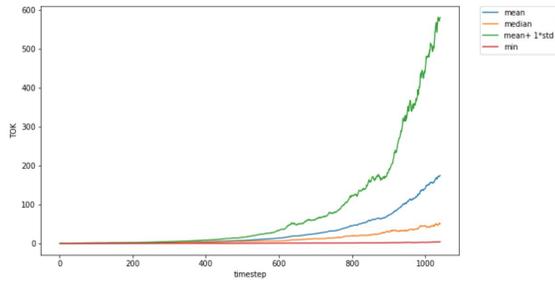
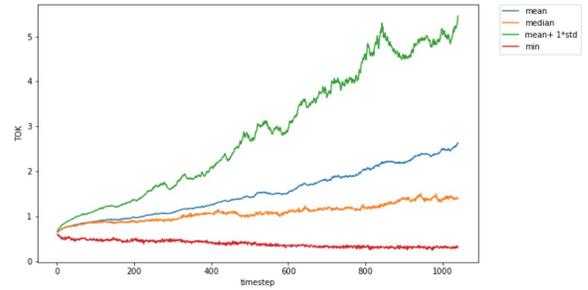
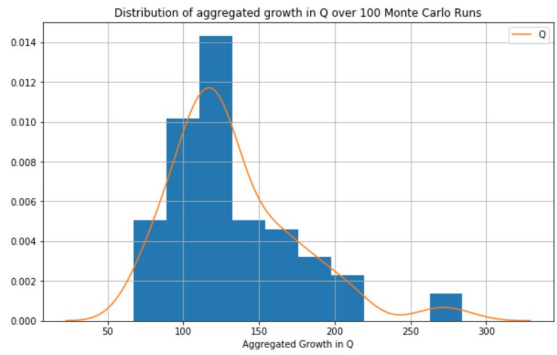
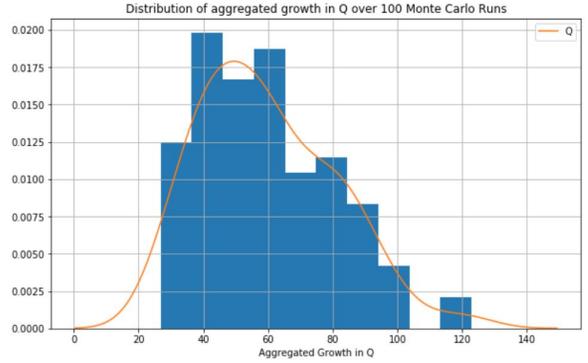
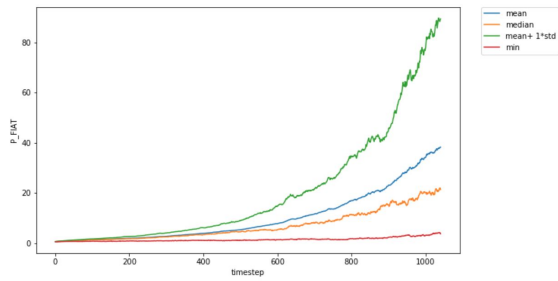
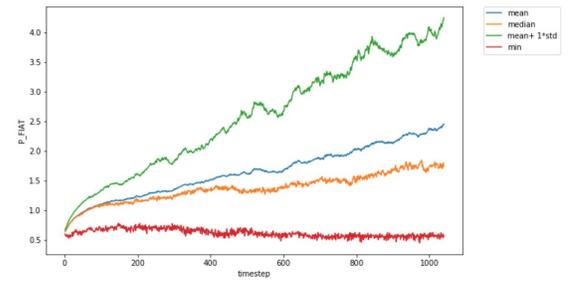
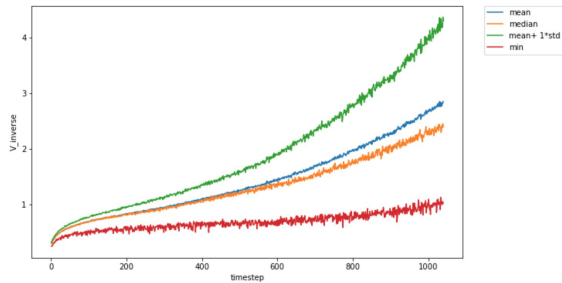
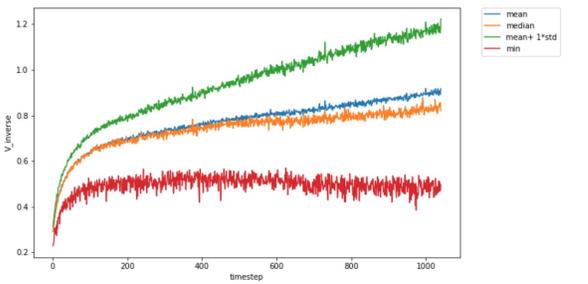

(a)                                              (b)

Figure 12. (a) Summary charts of high speculative value scenario
(b) summary charts of low speculative value scenario.

As we can see from Figure 12(a), a greater upward drift results in much higher TOK prices and a rightward shift in the distribution of aggregated growth in **Q(t)** with a mean of 133x. However, it also results in a much higher unit service price in fiat denomination with a variance of 283.96 and a mean of 10.24. This is very much the case in Bitcoin when the transaction fee in fiat terms became too high as the price of Bitcoin went above $10,000. In contrast, in Figure 12(b) where there is not much speculation going on in TOK prices. We observe a leftward shift in the distribution of aggregated growth in **Q(t)** with a mean of 60x and much lower TOK prices. Nonetheless, the unit service price is much lower in fiat currency with a smaller variance of 0.176. Note also that the intrinsic value of the token is also much higher with higher speculation, as speculation leads to a larger network with a higher level of activity as described in the positive feedback loop above.

As much as people criticized cryptocurrencies for their insane volatility, our simulation shows that consistent speculation can be a great way to seed a network. Fundamentally speaking, the existence of a token seeded the network with a community of stakeholders whose incentives are aligned to help the network grow and the token appreciate in value. Such an advantage is not present in most traditional ventures.

We have attempted to pull real world data from decentralized networks. However, the data quality is rather poor with no simple ways to ingest and query. In addition, the notion of a unit service that the network provides is not well defined. For instance, Ethereum is providing distributed computation as a service but the unit of computation is not well defined on the network level. This is important for token economies to scale beyond the circle of hobbyists to allow for proper business forecasting, demand projection, and large-scale adoption. Nonetheless, the assumption of a user-defined unit service enables us to abstract away the exact value of supply and demand and instead focuses on the system dynamics.

## Engineering Block Rewards

On a high level, we want a token economy where its native token continues to appreciate in intrinsic value, its service provided is highly demanded, and its service is transacted at a relatively stable price even in fiat denomination. The variable **B(t)**, the block rewards at time t, plays a very important role. A high **B(t)** represents a high subsidy that the network provides to subsidize its formation. However, it is also an inflationary force that dilutes the intrinsic value of the token. Nonetheless, **B(t)** is a big part of miner profitability **R(t)** that will impact the service supply. Hence, we want to engineer **B(t)** such that it gives out more block reward when the intrinsic value is decreasing. We now attempt to engineer a series of block rewards that keep track of the system performance in **1/V(t)** with a series of **B(t)** that follows the following update equation,

$$B(t) \sim \frac{V(t)}{V(t-1)} \times \frac{B(t-1)}{W(t-1)} \times W(t)$$

In other words, **B(t)** as a fraction of the remaining block rewards, **W(t)**, increases when the intrinsic value **1/V(t)** decreases and hence maintaining a driving force on **1/V(t)**. Given our new block rewards function, we will now perform 100 Monte Carlo runs with the same initial block rewards and speculative level as the baseline model.

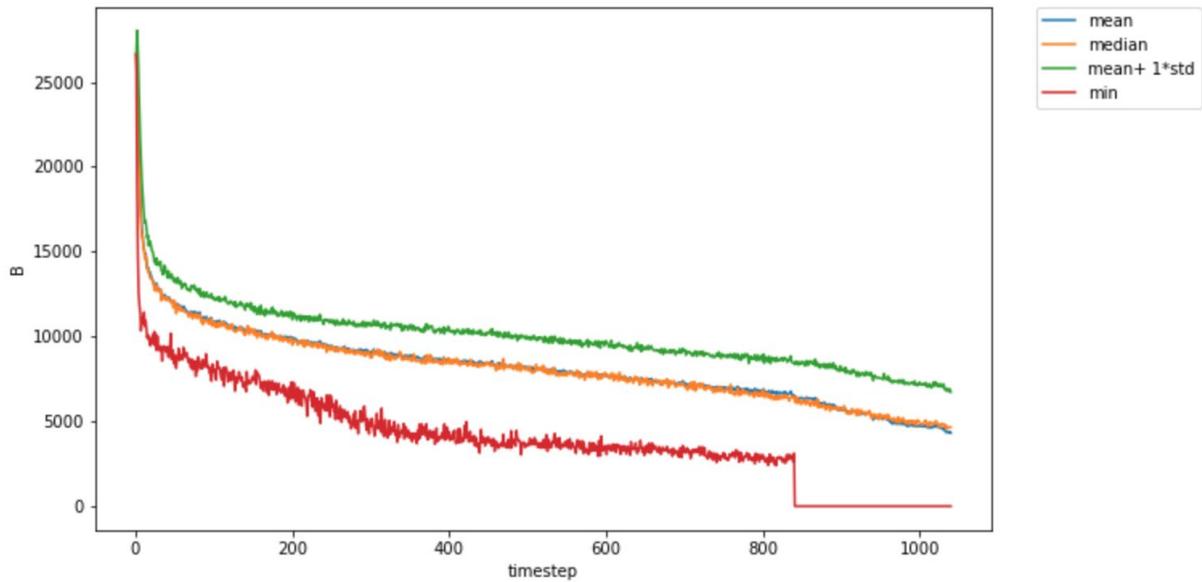

Figure 13. Block rewards over 100 Monte Carlo runs for V_inverse targeted model.

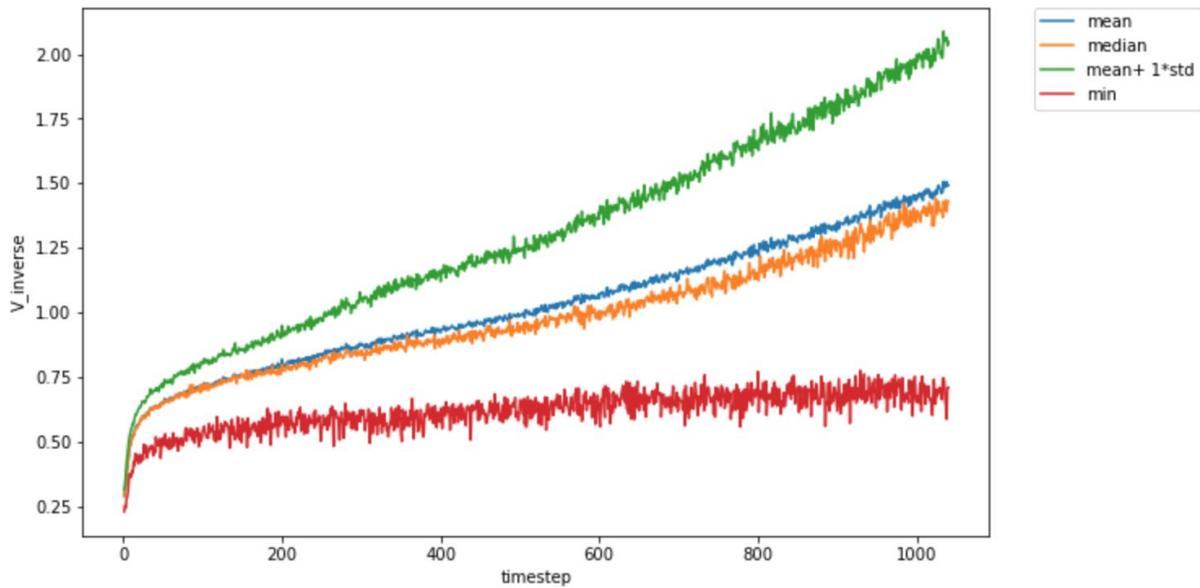

Figure 14. Summary of 1/V(t) over 100 Monte Carlo runs for V_inverse targeted simulation.

Note that in Figure 13, there is a sharp drop to zero for block rewards. That occurs when the allocated block rewards reserve **W(t)** gets depleted. In Figure 14, the terminal intrinsic value did not significantly outperform that in our baseline model. This can be attributed to two factors. One, the block rewards might have been depleted in later stages and its impact on value has reduced. Two, as **Q(t)** grows, the intrinsic value is largely dominated by **C(t)/P(t)**. However, it is worth noting that **1/V(t)** rose very sharply early in the network. The following table shows the time taken for the token value to reach certain milestones.

| V_inverse Value | 0.5 | 0.75 | 1.00 |
|---|---|---|---|
| Baseline Model | 28 weeks | 177 weeks | 529 weeks |
| V_inverse targeted Model | 9 weeks | 125 weeks | 504 weeks |

Table 2. Time taken for network to reach V_inverse milestones.

After 100 Monte Carlo runs on each of our models, it takes about one third of the time for the intrinsic value to hit 0.5 and one year in advance for the metric to hit 0.75. As the network progresses, block rewards get depleted and **Q(t)** becomes the dominant factor and hence the advantage of a targeted model becomes less obvious. A more accurate model of intrinsic value will have to capture the network effect value of the system but that is beyond the scope of the thesis.

We then attempted to run experiments that target the growth of **R(t)** in a similar way but given the volatility of **TOK(t)** as an input to **R(t)**, block rewards get depleted quickly and fail to lead to interesting results.

# Future Directions

A reasonable, robust, and informative model of the token economy can serve as the springboard to answer more important questions that are of interest to users, miners, investors, and protocol designers. The mathematical constructs developed and examined in this paper pave the way for more complex system and network analyses. In classical control engineering language, we are designing a useful model of a plant in this paper, that is a representation of a token economy in which the network pays miners in tokens to provide for a service of interest to consumers. We will now discuss future directions that our thesis can lead to and outline their respective challenges.

## Network Effect as a Positive Externality

Network effect can be loosely defined as the positive externality of a user to a platform economy when the user joins the network. Given the presence of network effect, the value of a product or platform increases with every additional user. Metcalfe's Law states that the value of a telecommunication network is proportional to the square of the number of users in the network. With our current set up, we consider the intrinsic value of the token as how much fiat money people are willing to put in exchange for the token. However, we do not capture the network effect that these token economy may achieve when it reaches a certain scale. After all, the holy grail of the most wildly successful ventures and platform economies is no doubt network effect. We would even argue that the ultimate goal of block reward subsidies is to bootstrap the network to a state in which strong network effect has been achieved.

Network effects in the context of traditional venture backed startups are straightforward to understand. It is also quite similar in token economies and cryptocurrencies. Every additional person believing in the value of Bitcoin will make Bitcoin more valuable because other users can now exchange goods and services with this person in Bitcoin. Similarly for Ethereum as another case in point, every additional user of decentralized applications will make Ethereum as the underlying computation platform and token more valuable.

The general heuristic for network effect is that users create more users, consumptions create consumptions, and capacity creates more capacity. We can potentially define an awareness vector that captures the awareness, sentiment, and perception of the network as a hidden state of the system. Changes in token price, user activity, media announcements, and so on will create a percentage change to this vector and the awareness vector will result in a faster adoption of the network. However, the modeling of network effect is another model in and of itself and hence we are not covering it in this thesis.

## Lyapunov Argument with Energy Field

Lyapunov argument is often used to provide guarantee on system stability around a subspace of attraction [32, 33]. If there is a maximum (or minimum) in an energy field and system energy is guaranteed to go up (or down) by at least a constant at every timestep, the system is bound to converge to the maximum (or the minimum). We have attempted to apply some of this concept into our system modeling. However, this will involve coming up with new energy field that may lose generality and breadth of application for our model since Lyapunov argument often only applies on a case-by-case basis.

The general stability guarantee provided by Lyapunov arguments means that in a closed-loop system, the system converges to a low energy state. An additional layer to that is input-output stability which is more robust in the sense that a system is only stable to the extent that a threshold amount of energy is injected into the system. Otherwise, the system remains stable and converges to a ball in space. The only way to move up to a higher energy level is through agents injecting energy into the system. How much energy injection is required to move the system into a certain energy state depends on the energy function of the system, which can be made to be a Lyapunov one. For someone to attack a cryptoeconomic system, Lyapunov stability enables us to encode the energy of the system and compute how much resources in fiat currency unit are required to move the system out of its stability. For instance, we can compute the amount of capital expenditure required to increase the energy of the system by 1% and the system can be designed to be insensitive to massive energy injection from agents. Such an argument is more robust than the economic stability arguments that are currently made in the decentralized economy ecosystem which often only focus on the cost to attack a system at some point in time without taking the temporal dynamics of the system into account.

## Optimal Agent Strategy

Another interesting angle to further this research is through the lens of an agent. Given the states and dynamics of the system, we hope to understand and compute the optimal strategies for a population of agents. We can define the following objective function for each agent i. Note that each agent i can also represent a subpopulation of agents with a similar profile.

$$max \, J_i = U_i(\sigma_i, X) - L(S_i, D_i, X)$$

where $U_i$ is a concave personal utility function of agent i with $\sigma_i$, agent's private beliefs, and X, the global observable system states, as inputs. Given the actions taken by the agent, $S_i$, supply side actions, and $D_i$, demand side actions, and the global state X, a cost is incurred as determined by the cost function L. This formulation is heavily inspired by literature in optimal control theory with infinite time horizon

[34]. As a control problem on infinite time horizon, discount on future utility has been accounted for by the utility function. Note that if one can properly define a terminating condition, such as achieving some definition of network effect, the above objective can be transformed to one that heavily rewards terminal utility and penalizes for the path taken to get there. It then becomes very similar to the strategy and optimization suggested in differential game literature [2].

# Conclusion

This thesis presents a methodology to model and simulate a generalized token economy in which the platform pays miners for a unit service provided in cryptocurrency. The market is assumed to be perfectly competitive with miners and users transacting a unit commodity service. To simulate the evolution of the system, we have defined a stochastic dynamical system and run Monte Carlo simulations to observe for system level behaviors and properties. Modifications to the model has been made to understand speculative price influence and explore the possibility of engineering block rewards for a specific property. Contrary to popular belief, token speculation on secondary market turns out to be beneficial to the network growth as it kicks start a positive feedback loop in adoption. Engineering block rewards to track token value also turns out to be viable but its impact will be better understood if we can include network value in modeling the value of the token. This thesis has been the first attempt to turn high level business narratives in cryptocurrency into observable and controllable simulation and experiment. We presented a framework and study on a generalized token economy that can be applied and extended easily to other more complex systems, turning the design of token economies from an art to a science.